\documentclass[conference]{IEEEtran}

\IEEEoverridecommandlockouts
\usepackage{cite}
\usepackage{amsmath,amssymb,amsfonts}
\usepackage{graphicx}
\usepackage{textcomp}
\usepackage{xcolor}
\usepackage[left=0.625in,right=0.625in,bottom=0.90in,top=0.9in]{geometry}
\usepackage{comment}

\usepackage[detect-all=true]{siunitx}
\DeclareSIUnit\decibelm{dBm}
\DeclareSIUnit\packets{packets}
\DeclareSIUnit\pct{percentile}
\AtBeginDocument{
	\DeclareSIUnit\bit{b}
	\DeclareSIUnit\bitpersec{bps}

	\newcolumntype{C}[1]{>{\centering\let\newline\\\arraybackslash\hspace{0pt}}m{#1}}
	\newcolumntype{L}[1]{>{\raggedright\let\newline\\\arraybackslash\hspace{0pt}}m{#1}}
	
}

\usepackage{soul}

\usepackage{booktabs}

\usepackage{cite}
\usepackage{amsmath,amssymb,amsfonts}
\usepackage{graphicx}
\usepackage{textcomp}
\usepackage{xcolor}

\usepackage{lipsum,multicol}
\usepackage{enumitem}
\usepackage{algorithm}
\usepackage {algpseudocode}
\usepackage{algorithmicx}
\usepackage[algo2e,ruled,vlined]{algorithm2e}

\usepackage{algcompatible}
\usepackage{subfigure}
\usepackage[subfigure]{tocloft}
\usepackage[bottom]{footmisc}
\usepackage{multirow}
\usepackage{amsthm,amssymb,amsmath,bm}
\usepackage{rotating}
\usepackage{empheq}
\usepackage{booktabs}
\usepackage{notoccite}
\usepackage{amsmath,cases}
\usepackage{pdfpages}
\usepackage{etoolbox}
\usepackage{lastpage}
\usepackage{fancyhdr}
\usepackage{framed}
\usepackage{float}
\usepackage{subfigure}
\usepackage{comment}
\usepackage{tikz}
\usetikzlibrary{spy}
\usepackage{authblk}
\usepackage[normalem]{ulem}
\usepackage{soul}

\begin{document}

\bstctlcite{IEEEexample:BSTcontrol}

\title{PDU-set Scheduling Algorithm for XR Traffic in Multi-Service 5G-Advanced Networks}



\author[1]{Pouria Paymard}
\author[2]{Stefano Paris}
\author[2]{Abolfazl Amiri}
\author[2]{Troels E. Kolding}
\author[2]{Fernando Sanchez Moya}
\author[1, 2]{\\Klaus I. Pedersen}
\affil[1]{Department of Electronic Systems, Aalborg University, Aalborg, Denmark}
\affil[2]{Nokia}
\maketitle
    
\begin{abstract}
    In this paper, we investigate a dynamic packet scheduling algorithm designed to enhance the eXtended Reality (XR) capacity of fifth-generation (5G)-Advanced networks with multiple cells, multiple users, and multiple services. The scheduler exploits the newly defined protocol data unit (PDU)-set information for XR traffic flows to enhance its quality-of-service awareness. To evaluate the performance of the proposed solution, advanced dynamic system-level simulations are conducted. The findings reveal that the proposed scheduler offers a notable improvement in increasing XR capacity up to 45\%, while keeping the same enhanced mobile broadband (eMBB) cell throughput as compared to the well-known baseline schedulers.
\end{abstract}

\section{Introduction}\label{P5:Introduction}
    
    The growing importance of eXtended reality (XR) services in the fifth-generation (5G)-Advanced standards communities \cite{3gpp.38.838, 3gpp.38.835}, and across various industries, calls for additional research to fulfill the required quality of service (QoS) targets for such services.
    In traditional mobile networks, QoS is primarily defined for flows and packets. Packets are scheduled by the radio access network (RAN) based on QoS requirements, such as the priority, guaranteed bit rate, user channel conditions, tolerated delay threshold, and so forth \cite{5GB_PS_survey}. Scheduling is a key mechanism to optimize network capacity while ensuring end-user satisfaction and is an intensively researched topic. As examples, \cite{5GB_PS_survey, LTE_PS_survey} provide comprehensive surveys on the schedulers widely used in the literature. The work in \cite{5GB_PS_survey} presents an overview of the state-of-the-art methods and investigates the current state of 5G radio resource management, while \cite{LTE_PS_survey} provides an overview of the key issues that arise in the design of a resource allocation algorithm for long-term evolution (LTE) networks. The authors list several schedulers and classify them for different use cases. A common way to handle the trade-off between fairness and spectral efficiency for the best-effort traffic is the use of proportional fair (PF) scheduling \cite{PF}. The authors of \cite{WPF} present the weighted PF (WPF) scheduling scheme to fulfill the flexible resource allocation considering the heterogeneous traffic pattern and QoS priority among users in a multi-service network. In the context of QoS provisioning, and especially when dealing with guaranteed delay requirements, the modified-largest weight delay first (M-LWDF) scheduler is widely used in the literature \cite{M-LWDF}. M-LWDF uses information about the head-of-line (HoL) delay for shaping the behavior of PF, assuring a good balance among spectral efficiency, fairness, and QoS provisioning \cite{M-LWDF}.
    
    While QoS-based user scheduling effectively works for certain mobile applications, it does not consider the inter-dependencies between packets. For instance, in the case of delivering a 4K video stream, each video frame may require around 1000 internet protocol (IP) packets with a combined size larger than 1 MB. If any of these packets fail to reach the application layer, the user experience is affected \cite{Standard_Evolution_5GA, frame_integration_Huawei}. Recently, the work in \cite{frame_integration_Huawei} proposes a heuristic scheduling algorithm that captures inter-dependencies between packets to improve the XR performance. For this, the scheduling metric includes the fraction of the successfully transmitted packets of the big XR video frame. 
    
    To enable the integrated transmission of video frames, the concept of a protocol data unit (PDU)-set is introduced in 5G-Advanced \cite{3gpp.23.700-60, RP-223502}. A PDU-set represents a collection of packets that carry a single media unit, e.g., a video frame. 5G-Advanced can leverage PDU-set related information, including PDU-set sequence number, PDU-set importance, PDU-set delay budget (PSDB), PDU-set error rate, etc., to support QoS handling at the PDU-set level \cite{3gpp.38.835}.
        
    This paper introduces advancements in packet scheduling techniques, surpassing previously published studies. The scheduling decisions explicitly consider the buffering time of PDU-set payload rather than solely focusing on individual PDU delays. Our scheduler also incorporates PDU-set segmentation awareness by prioritizing the scheduling of the PDUs of a PDU-set of an XR user equipment (UE) in consecutive transmissions. We aim to obtain highly realistic system-level performance results for the co-existence of XR and enhanced mobile broadband (eMBB) UEs in a dynamic multi-cell 5G New Radio (NR)-compliant system. To achieve this, we choose advanced system-level simulations (SLSs) as our primary methodology for performance analysis. The SLS relies on a comprehensive set of mathematical models that have been endorsed by the 3rd Generation Partnership Project (3GPP) standardization and widely adopted in academia.

\section{setting the scene}\label{P5:System_Model}
    
\subsection{Traffic Model}\label{P5:Traffic_Model}
\begin{figure}[t]
        \centerline{\includegraphics[width=0.37\textwidth,trim={0cm 0cm 0cm 0cm},clip]{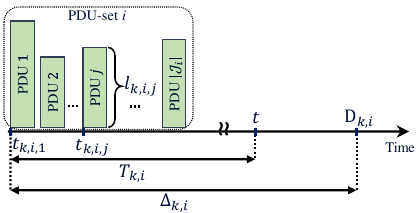}}
        \caption{The relation between PDUs within a PDU-set.}
        \label{P5:Fig:PDU-set}
    \end{figure}

    We assume a mixed traffic scenario where both XR and eMBB users coexist.
    We assume that an XR server in the proximity of the next-generation nodeB (gNBs) generates XR video frames with a fixed frame rate of 60 frames per second (fps). This gives a periodicity of \SI{16.67}{ms}. A random delay jitter impacts the exact periodicity of frame arrival time at the gNB. The jitter follows a truncated Gaussian distribution $\mathcal{TN}(\mu_m, \sigma_m, a_m, b_m)$, where $\mu_m$ and $\sigma_m$ express the mean and the standard deviation, and it truncates within the interval of $(a_m, b_m)$. A PDU-set is defined as a collection of PDUs that carry one XR video frame. The PDU-set size is also assumed to follow a truncated Gaussian distribution, denoted as $\mathcal{TN}(\mu_f, \sigma_f, a_f, b_f)$, with $\mu_f$ representing the mean, $\sigma_f$ indicating the standard deviation, and non-zero interval of $(a_f, b_f)$. We express the XR UE, PDU-set, and PDU indices with $k\in\mathcal{K}$, $i\in\mathcal{I}_k$, and $j\in\mathcal{J}_i$, respectively. Figure \ref{P5:Fig:PDU-set} exemplifies the relation between the PDU-set and PDUs.
    The size of each PDU $j$ of PDU-set $i$ of UE $k$ is assumed to be $l_{k,i,j}$.
    PDU $j$ of PDU-set $i$ of UE $k$ arrives at $t_{k,i,j}$ to the gNB buffer, while the first PDU of PDU-set $i$ arrives at the gNB buffer in $t_{k,i,1}$. The HoL delay of PDU-set $i$ of UE $k$ is defined as $T_{k,i}=t-t_{k,i,1}$, where $t$ is the current time. The deadline for transmitting the whole PDU-set $i$ is represented by $D_{k,i}$ which is equal to $\Delta_{k,i}+t_{k,i,1}$, where $\Delta_{k,i}$ is the PSDB in ms. 
    
    We represent the eMBB UEs with $e\in\mathcal{E}$. For the eMBB traffic model, we assume the best-effort full buffer model. This implies that there is always data buffered at the gNB, waiting to be transmitted to eMBB UEs. Consequently, in cells where eMBB UEs are present, the cell operates at full load consistently, as there are infinite pending data at the gNB to utilize all available radio resources for transmission.
    In line with the 3GPP evaluation methodology \cite{3gpp.38.838, 3gpp.38.835}, we define the main key performance indicators (KPIs) for XR studies. An XR UE is labeled satisfied if at least 99\% of its PDU-sets are successfully received within the given PSDB. The UE satisfaction ratio is calculated as the ratio of the number of satisfied UEs divided by the total number of XR UEs present in the network. XR capacity is defined as the maximum number of XR UEs per cell, with at least 90\% of them being satisfied. 
    For the eMBB UEs, we assume the average eMBB cell throughput (TP) as our primary KPI.

\subsection{Deployment Model}
    We adopt the indoor hotspot (InH) deployment scenario in a 5G-Advanced NR network detailed in \cite{3gpp.38.901}, with a focus on the downlink performance in time division duplexing (TDD) mode. The TDD mode follows a DDDSU slot pattern, where D, S, and U denote downlink, special, and uplink slots, respectively. Our network configuration comprises 12 cells, forming a 2-row deployment of gNBs with an inter-site distance of 20 meters. Multiple XR and eMBB UEs are randomly distributed across the entire network, with an equal number of UEs per traffic type connected to each gNB. 

    A 100MHz channel bandwidth is shared using orthogonal frequency division multiple access (OFDMA). Sub-carrier spacing of 30 kHz is assumed, resulting in a slot duration of \SI{0.5}{ms} and a total of 272 physical resource blocks (PRBs) \cite{3gpp.38.211}. A slot and a PRB are represented by $s\in\mathcal{S}$ and $p\in\mathcal{P}$, respectively.
    
    Dynamic scheduling and link adaptation are assumed in line with the assumptions in \cite{JSAC_eCQI, GlobeCom_eOLLA}. This includes code block group (CBG) based transmissions with multi-bit hybrid automatic repeat request (HARQ) feedback. If a transmission is not correctly decoded, the UE provides feedback to the gNB to only retransmit the CBGs not correctly received by the UE. Enhanced channel quality indicator (CQI) feedback \cite{JSAC_eCQI} and gNB outer loop link adaptation tailored for CBG-based cases are adopted \cite{GlobeCom_eOLLA}. The HARQ is asynchronous with Chase combining.

\section{Proposed PDU-set Scheduling}\label{P5:PDU-Set_Scheduling}
\subsection{Problem Formulation}\label{P5:Problem_Formulation}
    In a deployment with mixed traffic where both XR and eMBB UEs coexist, the scheduler allocates radio resources over the time window $\vert S\vert$ to maximize the number of satisfied XR UEs and provide a fair share of radio resources to eMBB UEs at the same time. This objective is formulated as follows:
    \begin{align} \label{pss:obj_function}
        &\max\limits_{\boldsymbol{\gamma, x,y,z}}
        ~~ \sum_{k\in\mathcal{K}} a_{k} . \gamma_{k} + \sum_{e\in \mathcal{E}} \log (R_e^{\text{eMBB}}),
    \end{align}
    where $\gamma_{k}$ indicates whether XR UE $k$ is satisfied. $a_{k}$ is a constant QoS parameter to show a higher priority for XR UEs than eMBB UEs.
    $R_{e}=\sum_{s\in\mathcal{S}}\sum_{p\in\mathcal{P}} x_{e,s,p} R_{e,s,p}$ is the average TP of eMBB UE $e\in\mathcal{E}$ measured over the time window $\vert S\vert$ of the scheduling decision. 
    The data rate denoted as $R_{e,s,p}$ for UE $e$, which can be transmitted over PRB $p$ on slot $s$, depends on the modulation and coding scheme (MCS) and the multiple-input and multiple-output (MIMO) rank selected by the gNB.
    The binary decision variable $x_{e,s,p}\in\{0,1\}$ is defined to denote whether the PRB $p$ in slot $s$ is allocated to UE $e\in\mathcal{E}$.
    
    As discussed in previous sections, XR UE is satisfied (i.e., $\gamma_{k}=1$) when 99\% of the PDU-sets are successfully served within a certain delay budget. To mark a UE as satisfied, we define the indicator variable $y_{ki}$ for each PDU-set $i\in\mathcal{I}_k$ of every UE $k$ and the following constraint, where $\gamma_{k}=1$ in the right-hand-side of the inequality only if 99\% of the PDU-sets in $\mathcal{I}_k$ have been scheduled:
    \begin{align} \label{pss:fraction_pdusets_within_psdb}
    &\sum_{i\in \mathcal{I}_k} y_{k,i} \geq \gamma_{k} \lceil \vert \mathcal{I}_{k} \vert \cdot 0.99 \rceil,  \quad\, \forall k\in\mathcal{K}.
    \end{align}
    
    We observe that a PDU-set $i\in\mathcal{I}_k$ of an XR UE $k$ is successfully delivered if enough resources are allocated to transmit the whole data within the delay budget. In other words, the amount of bits of the radio resources allocated to the PDU-set must cover its size:
    \begin{align} \label{pss:tot_prbs_per_ue}
    &\sum_{s\in \mathcal{S}} \sum_{p\in \mathcal{P}} x_{k,s,p} . L_{k,s,p} \geq \sum_{j\in \mathcal{J}_i} y_{k,i} . l_{k,i,j}  , \,\forall k\in\mathcal{K}, i \in \mathcal{I}_k,
    \end{align}
    where, $L_{k,s,p}$ is data bits for UE $k\in\mathcal{K}$ transmitted over PRB $p$ on slot $s$.
    Additionally, the last time slot $s$ used to serve the PDU-set must be smaller or equal than the delay budget. To this end, we first define the indicator variable $z_{k,i,s}$ to identify the slots that are used to serve each PDU-set $i$ of each UE $k$ in~\eqref{pss:slot_index} and we limit the allocation to those slots that do not exceed the PDU-set delay budget $D_{k,i}$ in~\eqref{pss:delay_budget}:
    \begin{align} 
    &\label{pss:slot_index}
    \sum_{p\in \mathcal{P}} x_{k,s,p} \leq z_{k,i,s} . |P|,  \quad\,\forall k\in\mathcal{K}, s \in \mathcal{S}, \\
    &\label{pss:delay_budget}
    z_{k,i,s} s \leq D_{k,i}, \quad\,\forall k\in\mathcal{K}, i\in\mathcal{I}_k, s \in \mathcal{S}.
    \end{align}
    
    Finally, in a single-user MIMO (SU-MIMO) system, the scheduler can allocate a PRB in any slot to a single UE, being XR or eMBB:
    \begin{align} 
    &\label{pss:one_ue_per_prb}
    \sum_{k\in\mathcal{K}} x_{k,s,p} + \sum_{e\in \mathcal{E}} x_{e,s,p}  \leq 1,  \forall s \in \mathcal{S}, p \in \mathcal{P}.
    \end{align}
    
    The PDU-set scheduling problem can be formulated as the following mathematical problem:
    \allowdisplaybreaks
    \begin{subequations}\label{Optimization_Problem}
    \begin{align} 
    &\max\limits_{\boldsymbol{\gamma, x,y,z}}
        ~~ \sum_{k\in\mathcal{K}} a_{k} . \gamma_{k} + \sum_{e\in \mathcal{E}} \log (R_e^{\text{eMBB}}),
    \\
    &
    \textrm{s.t.} ~
    \eqref{pss:fraction_pdusets_within_psdb}-\eqref{pss:one_ue_per_prb} \nonumber
    \\
    &\label{pss:var_allocation}
    x_{k,s,p}, x_{e,s,p} \in\lbrace 0,1 \rbrace, k\in\mathcal{K}, e\in\mathcal{E}, s \in \mathcal{S}, p \in \mathcal{P}, \\
    &\label{pss:var_admission}
    y_{k,i}\in\lbrace0,1\rbrace,  \forall k\in\mathcal{K}, i \in \mathcal{I}_k, \\
    &\label{pss:var_slot}
    z_{k,i,s}\in\lbrace0,1\rbrace,  \forall k\in\mathcal{K}, i \in \mathcal{I}_k, s \in \mathcal{S}.
    \end{align}
    \end{subequations}

\subsection{Low-complexity Proposed Solution}\label{P5:Proposed_Scheduler}
    We observe that the problem \eqref{Optimization_Problem} is a bin-covering problem, which is known to be NP-hard \cite{NP_Hard}. The optimization problem  \eqref{Optimization_Problem} can be solved through a brute-force algorithm, but it comes with a computational complexity of $\mathcal{O}\left(\left|P\right|^{B\times\left(\left|U\right|+\left|E\right|\right)}\right)$, where $B$ represents the number of cells in the network, while $|P|$, $|U|$, and $|E|$ denote the cardinality of the sets $\mathcal{P}$, $\mathcal{K}$ and $\mathcal{E}$, respectively. Consequently, this high-complexity solution may not be feasible for real wireless networks since scheduling decisions must be made rapidly, typically on a per-transmission time interval (TTI) basis. 

    Instead, a heuristic solution is proposed to allow the problem to be solved in real-time. 
    The solution comprises a new scheduling policy as well as the design of a new scheduling metric including the newly defined PDU-set information. 
    on each TTI, each cell assigns PRBs to their respective users taking into account the user-specific QoS class, the current buffer status of UEs, and XR-specific PDU-set information including PSDB, PDU-set size, and served bytes of PDU-set.
    Considering this, we introduce two key parameters to the scheduling metric: 
    \begin{enumerate}
        \item $\alpha_{k,i}$ represents the ratio of transmitted bits of PDU-set $i$ of UE $k$ and is expressed as follows:
        \begin{equation}
            \alpha_{k,i}=\frac{w_{k,i}}{\sum_{j\in\mathcal{J}_i}\ l_{k,i,j}}, 
        \end{equation}
        where $w_{k,i}$ and $\sum_{j\in\mathcal{J}_i}\ l_{k,i,j}$ represent the successfully transmitted bits and the size of PDU-set $i$ of UE $k$, respectively.
        
        \item $\beta_{k,i}$ represents the ratio of the remaining delay budget of PDU-set $i$ of UE $k$ and is expressed as follows:
        \begin{equation}
            \beta_{k,i}=1-\frac{T_{k,i}}{\mathrm{\Delta}_{k,i}}. 
        \end{equation}
    \end{enumerate}
    The inclusion of $\alpha_{k,i}$ in the scheduling metric prioritizes the scheduling of the entire PDU-set of an XR UE across consecutive transmissions. Moreover, $\beta_{k,i}$ accounts for the buffering time of PDU-set payload rather than solely focusing on individual PDU delay or PDU delay budget. 
    Note that each XR frame is rather large (in the range of 46kB-141kB) so it requires the full bandwidth to transmit a PDU set in a few TTIs. Secondly, given a large number of antenna elements at the gNB and UE, the effective channel has little frequency selectivity and no gains from frequency-selective channel-aware scheduling. 
    Consequently, we define our proposed scheduling metric for each UE $k$ and PDU-set $i$ as follows:
    \begin{equation}\label{P5:PS_Scheduler_Metric}
        m_{k,i}=\frac{e^{\alpha_{k,i}}}{\beta_{k,i}}\times u\left[\mathrm{\Delta}_{k,i}-T_{k,i}\right],
    \end{equation}
    where $u\left[\mathrm{\Delta}_{k,i}-T_{k,i}\right]$ expresses the unit step function to de-prioritize scheduling PDU-set $i$ of UE $k$ when its HOL delay $T_{k,i}$ exceeds PSDB $\mathrm{\Delta}_{k,i}$. As per 3GPP technical specification group service \& system aspects (TSG SA) studies, the delayed PDUs are still useful to be transmitted \cite{R2-2206969}. Consequently, the delayed PDUs are still scheduled but only towards the end, when there are no XR UEs with in-time PDUs to schedule.

    Algorithm \ref{P5:Alg:RRA} summarizes the proposed scheduling approach. The PRBs are allocated to the candidate UEs in the following manner: First, HARQ retransmissions are scheduled, utilizing the available PRBs. Pending HARQ retransmissions of XR UEs are prioritized over those of eMBB UEs. Next, the remaining PRBs are allocated to new transmissions. 
    To reduce the XR queuing delay and enhance XR reliability, the scheduler organizes the new transmission scheduling into two stages: 1) XR payloads are scheduled. 
    To achieve this, the XR UEs are ordered in descending order of priority metric as defined in \eqref{P5:PS_Scheduler_Metric}. The UE with the highest priority metric is granted the first scheduling opportunity.
    2) After scheduling all XR UEs, eMBB UEs are scheduled on the remaining PRBs according to the PF metric as follows:
    \begin{equation}\label{P5:PF_Metric}
        m_k=\frac{r_k}{{\bar{R}}_k},
    \end{equation}
    where $r_k$ and ${\bar{R}}_k$ are the instantaneous full-bandwidth TP, and the past achieved TP, respectively, for UE $k$. 
    The allocation process continues until either there are no more available PRBs or there are no more users left to schedule.
    \begin{algorithm}[t]
        \caption{Radio resource scheduling per TTI}
        \label{P5:Alg:RRA}
        \begin{algorithmic}[1]
            \STATEx \textbf{Inputs \& Parameters:}
            \STATEx $\mathbb{C}_{\text{UE}}$: All candidate UEs including a subset of XR ($\mathcal{K}$) and eMBB ($\mathcal{E}$) UEs waiting to be scheduled,
            \STATEx $M_{\text{UE}}$: Boolean parameter states whether the UE has data in the gNB buffer for transmission,
            \STATEx $\mathbb{V}_{\text{PRB}}$: Available PRBs in the TTI

            \medskip
            \STATEx \textbf{Output:} Pair of allocated PRBs to the corresponding UEs
            
            \medskip
            \STATE Schedule the HARQ retransmissions
            \STATE Update $\mathbb{V}_{\text{PRB}}$
            \STATE Order the set of candidate UEs based on their traffic type
            \STATE Sort XR and eMBB UEs based on scheduling metrics expressed in \eqref{P5:PS_Scheduler_Metric} and \eqref{P5:PF_Metric}.
            \FOR{UEs $\in \mathbb{C}_{\text{UE}}$}
            \WHILE{($M_{\text{UE}} \And |\mathbb{V}_{\text{PRB}}| > 0$)}
            \STATE Assign one PRB
            \STATE Update $\mathbb{V}_{\text{PRB}}$
            \ENDWHILE
            \ENDFOR
        \end{algorithmic}
    \end{algorithm}
     
                
                
    	 
                
            
            
    
    
            
            
            
    
\section{System-level Simulation (SLS) Results}\label{P5:SLS}

\subsection{Evaluation Methodology}\label{P5:Evaluation_Methodology}    
    We employ advanced SLSs which have been calibrated in accordance with the 3GPP simulation methodology described in \cite{3gpp.38.838, 3gpp.38.835}.
    The simulator models the radio access network user plane protocols, radio resource management (RRM) mechanisms, 3D radio propagation, SU-MIMO with rank adaptation, and traffic models commonly adopted for 3GPP evaluation \cite{3gpp.38.838, 3gpp.38.835}. Minimum mean square error interference rejection combining (MMSE-IRC) receivers at UEs are assumed \cite{MMSE_IRC}. 
    
    After each transmission, the effective signal-to-interference-plus-noise ratio (SINR) for each of the assigned resource elements (REs) is calculated, and then the mean mutual information per coded bit (MMIB) is computed over all the scheduled CBGs. The calculated MMIB value along with the knowledge of transmitted MCS is used to determine the error probability of transmitted CBGs from detailed look-up tables that are obtained from extensive link-level simulations \cite{tang2010mean}. 
    
    We evaluate the proposed scheduler against two well-known baseline scheduling algorithms: 
    \begin{enumerate}
        \item WPF \cite{WPF} with the scheduling metric for each UE $k$:
            \begin{equation}\label{P5:WPF_Metric}
                m_k=w_k\times\frac{r_k}{{\bar{R}}_k},
            \end{equation}
            where $w_k$ is the respective weight for UE $k$. Since XR has higher priority, the ratio of XR UEs’ weight to eMBB UEs’ is assumed to be $10^{8}$.
        
        \item M-LWDF \cite{M-LWDF} with the scheduling metric for each UE $k$ and PDU-set $i$ being expressed as:
            \begin{equation}\label{P5:MLWDF_Metric}
                m_{k,i}= -\log{\left(\delta\right)} \times\frac{T_{k,i}}{\mathrm{\Delta}_{k,i}}\times\frac{r_k}{{\bar{R}}_k},
            \end{equation}
            where 
            the term $-\log{\left(\delta\right)}$ 
            performs a role analogous to $w_k$ for determining the prioritization of XR UEs over eMBB UEs. As there are no delay requirements for eMBB UEs, $\frac{T_{k,i}}{\mathrm{\Delta}_{k,i}}$ in \eqref{P5:MLWDF_Metric} is omitted for eMBB UEs.
    \end{enumerate}
    Both baseline schedulers are QoS- and radio channel-aware. The M-LWDF variant also includes delay awareness.

    To ensure statistically reliable results, we conduct simulations to collect a minimum of 540 PDU-set transmissions for each XR UE. Considering uncorrelated samples, this enables us to estimate with a 99\% confidence level, if 99\% of the packets are correctly received within the PDB with an error margin of no more than $\pm 1\%$. This is the criterion for determining XR UE satisfaction in the 3GPP evaluation methodology. In each simulation, we assume $N$ XR UEs/cell and 12 cells, resulting in statistics collected from $12 \times N$ users. Additionally, each simulation consists of $M = 10$ drops, where the UEs are randomly redistributed in different positions. Therefore, our statistics include $12 \times N \times M$ XR users. The number of XR users per cell is varied from 3 to 8.
    For instance, when $M = 10$ and $N = 6$, we obtain statistics from 1200 XR users, which allows us to estimate, with a 99\% confidence level, if 90\% of the XR UEs are satisfied according to the XR capacity definition metric, with an error margin of at most $\pm2\%$. This approach is sufficient to monitor the TP samples per TTI and obtain accurate eMBB average cell TP statistics.

    \begin{table}[t]
		\centering
		\caption{Summary of System-level Evaluation Parameters}
		\begin{tabular}{c c}
			\textbf{Parameter} & \textbf{Setting} \\ \hline \hline 
			Deployment & InH \\ 
                World Area & 120m $\times$ 50m  \\
			Layout  & 12 cells  \\ 
			Inter-site Distance & 20 m \\ 
   			gNB height & 3 m  \\ 
			gNB Tx power & 31 dBm  \\ 
                gNB down tilt & $90^{\circ}$  \\ \hline
			TDD Frame structure & DDDSU\\ 
			TTI length & 14 OFDM symbols\\
			PDCCH overhead & 1 OFDM symbol\\
			Carrier frequency & 4 GHz\\ 
		      Bandwidth & 100 MHz\\ 
			Sub-carrier spacing & 30 kHz\\ \hline
			MIMO scheme & SU-MIMO with rank adaptation \\ 
			Modulation & QPSK to 256QAM\\
			gNB Tx processing delay & 2.75 OFDM symbols\\ 
                \multirow{2}*{gNB antenna} & 1 panel with 32 elements\\
			& (4 x 4 and 2 polarization)\\ \hline
			UE speed & 3 km/h \\ 
			UE height & 1.5 m \\ 
			UE Rx processing delay & 6 OFDM symbols\\
                UE receiver & MMSE-IRC\\
			UE antenna polarization & dual-polarized\\
			Number of UE antennas & 2 Rx antennas\\ \hline
                eMBB Traffic model & full-buffer \\ 
                number of eMBB UE/cell & 3 \\ 
			Random Jitter & $\mathcal{TN}(0, 2, -4, 4)$ ms\\
			XR frame rate & 60 fps\\ 
			XR frame size & $\mathcal{TN}(93, 10, 46, 141)$ kB\\ \hline
			HARQ scheme & CBG-based HARQ retransmissions\\ 
                HARQ combining method & Chase combining\\
                Channel estimation & Realistic\\
			CQI & Periodic CQI every \SI{2.5}{ms} \\
			Target CBG error probability & 2 out 8 CBGs with probability of 50\% \\  \toprule \toprule
		\end{tabular}
		\label{P5:Table:System_Parameters}
\end{table}

\subsection{Performance Results}\label{P5:Performance_Results} 
\begin{figure}[t]
        \centerline{\includegraphics[width=0.37\textwidth,trim={0cm 0cm 0cm 0cm},clip]{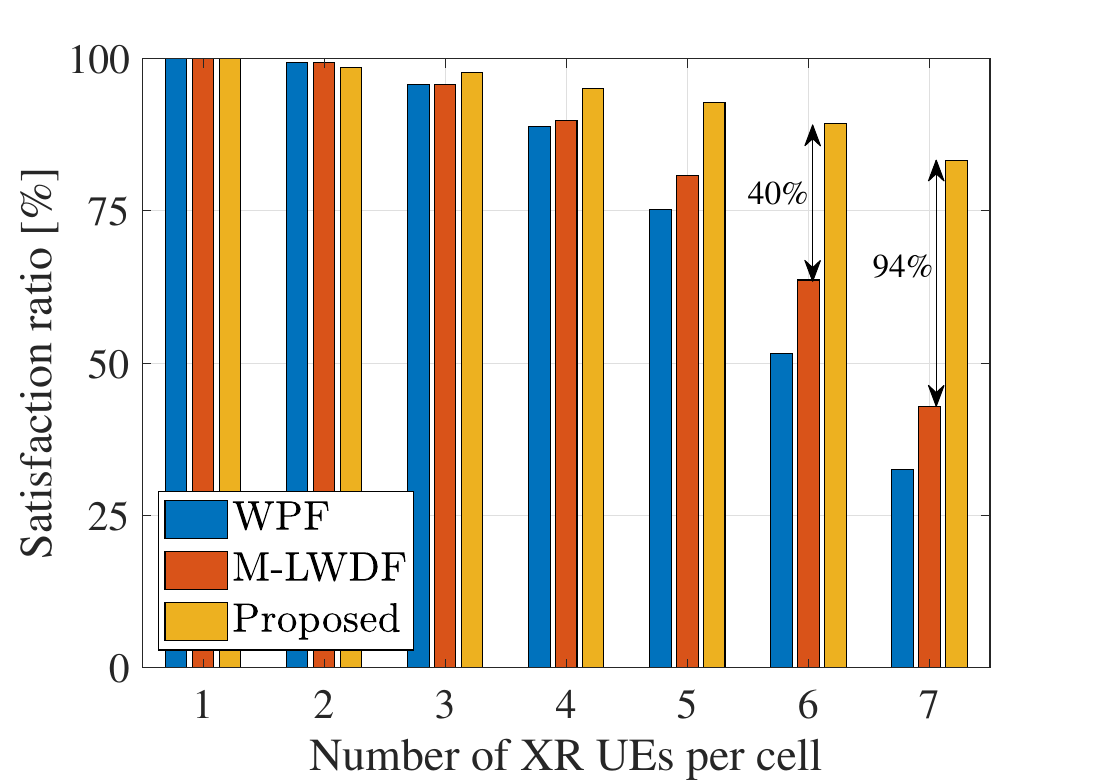}}
        \caption{XR Satisfaction ratio versus number of connected XR UEs per cell for three different schedulers.}
        \label{P5:Fig:SatisfactionRatio}
      \end{figure}

    Figure \ref{P5:Fig:SatisfactionRatio} shows the XR satisfaction ratio results versus the number of connected XR UEs per cell for three different schedulers when the PSDB is 15 ms. As expected, XR satisfaction ratio drops when adding more XR UEs to the network. The proposed scheduler ensures a smooth reduction in the XR satisfaction ratio in high-load scenarios, whereas the baseline schedulers experience a significant drop. For example, when the system load is 7 XR UEs per cell, the proposed scheduler enhances the XR satisfaction ratio by 94\% as compared to the M-LWDF scheduler. Overall, the proposed scheduler consistently outperforms the baseline schedulers in all scenarios due to its enhanced awareness of PDU-set segmentation.

\begin{figure}
        \centering
            \centering
            \includegraphics[width=0.37\textwidth,trim={0cm 0cm 0cm 0cm },clip] {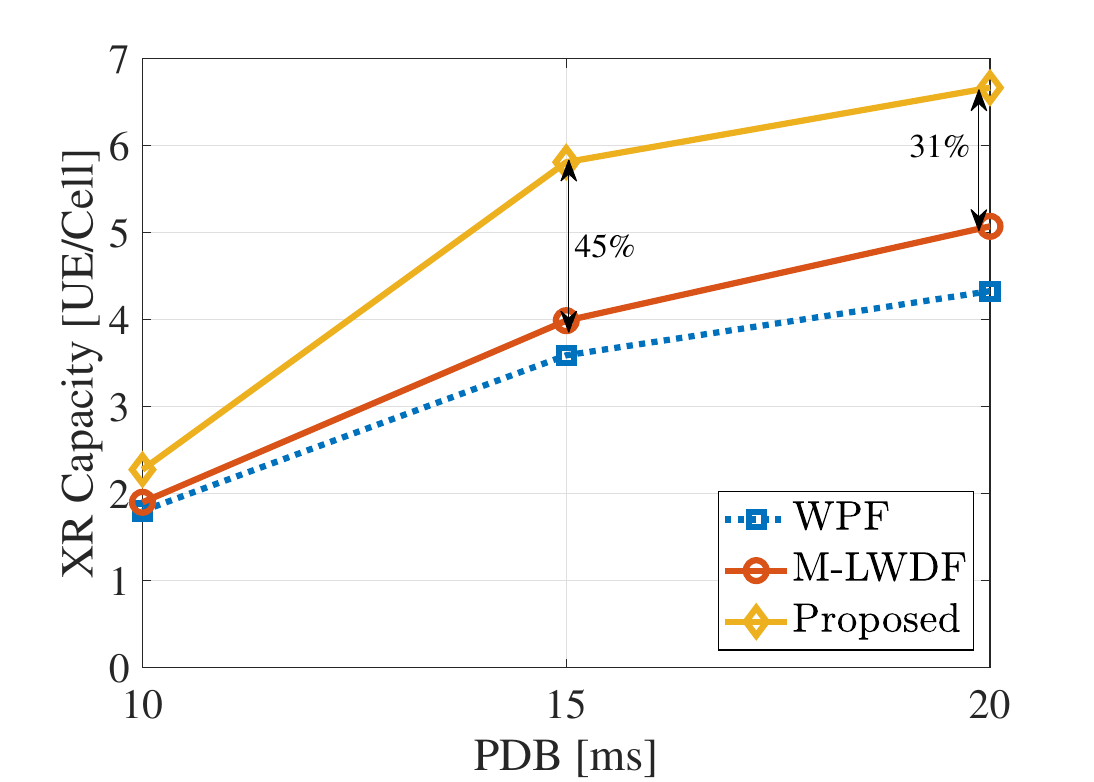}
        \caption{XR capacity versus PDB for three different schedulers.}
        \label{P5:Fig:XR_capacity_PDB}
    \end{figure}

    The PSDB value depends on the XR application. As examples, cloud gaming cases typically have PSDB values of 15 ms and 20 ms, while stricter PSDB, such as 10 ms, are assumed for VR cases \cite{3gpp.38.838, 3gpp.38.835}. Figure \ref{P5:Fig:XR_capacity_PDB} illustrates the influence of PSDB on the XR capacity. The results are presented for the three schedulers and PSDB values of \{10, 15, 20\} ms. As expected, when the PSDB is tightened, the XR capacity decreases to fewer UEs. When the PSDB is extremely tight ($\le$ \SI{10}{ms}), all three schedulers show minimal differences. The primary reason is that the scheduler has a reduced degree of freedom as it should schedule any incoming data immediately to fulfill the latency constraint. Hence, the scheduler does not have much of the freedom to slightly postpone scheduling some UEs to prioritize others, and vice versa. 
    In contrast, the proposed scheduler benefits from the larger flexibility offered by larger PSDB values and it outperforms the baseline schedulers.  
    As an example, with a PSDB of 15 ms, the proposed scheduler improves the XR capacity from 4 UEs per cell to 5.8 UEs per cell compared to the M-LWDF scheduler, corresponding to about a 45\% increase in the XR capacity.

\begin{figure}
        \centering
        \includegraphics[width=0.37\textwidth,trim={0 0 0 0},clip] {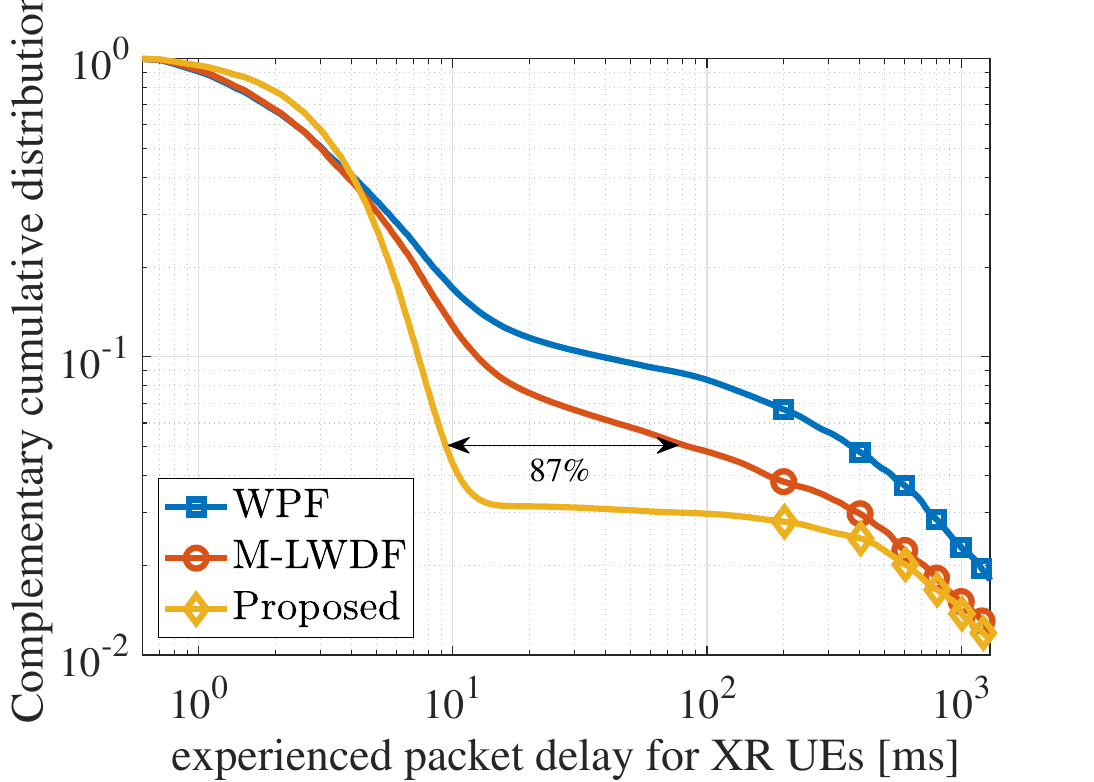} 
                 
        \caption{The empirical CCDF of experienced packet delay for XR UEs.}
        \label{P5:Fig:CCDF_of_Delay_45Mbps}
    \end{figure}

    Figure \ref{P5:Fig:CCDF_of_Delay_45Mbps} shows the empirical complementary cumulative distribution function (CCDF) of the experienced PDU-set delay for XR UEs when there are 6 XR and 3 eMBB UEs/cell. It consists of delays from when PDU-sets are generated at the XR server to when they are received correctly by the application layer. It accounts for various effects such as potential gNB queuing, processing times at the gNB and UE, potential HARQ retransmissions, etc. The CCDF is built on the collection of experienced delays from all the UEs in the network. As seen, the proposed scheduler significantly outperforms the baseline schedulers. As an example, the proposed scheduler reduces the delay from 80 ms to 10 ms and accordingly 87\% improvement at the 95\textsuperscript{th}-percentile of the CCDF. The notable advantage of the proposed scheduler lies in its knowledge of the PDU-set segmentation and in downgrading the transmission priority of delayed PDU-set payloads that exceed the PSDB. By scheduling PDUs belonging to a PDU-set across consecutive slots, the proposed scheduler effectively reduces the overall delay of the PDU-set, ensuring all the PDUs are received before the PSDB and hence the PDU-set is decodable.

\begin{figure}
        \centering
        \includegraphics[width=0.37\textwidth,trim={0 0 0 0},clip] {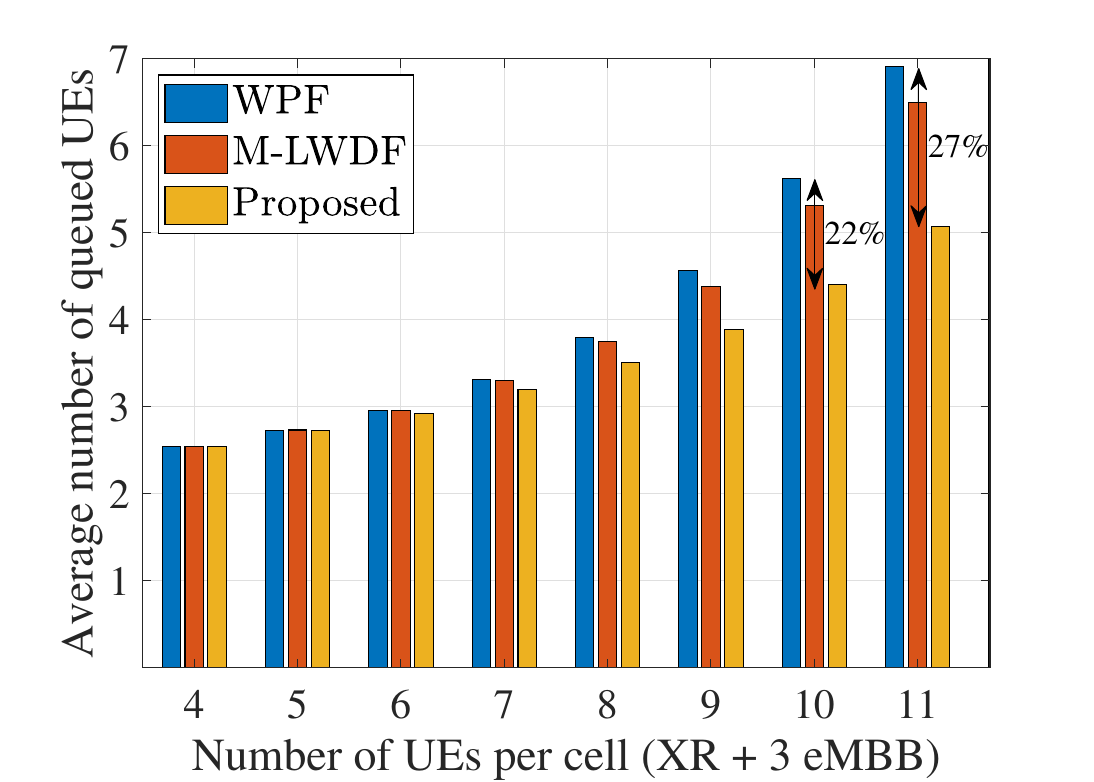}
                 
        \caption{Average number of queued UEs waiting to be scheduled versus connected XR UEs per cell for three different schedulers.}
        \label{P5:Fig:Queued_Users}
    \end{figure}

    Figure \ref{P5:Fig:Queued_Users} illustrates the relationship between the number of queued UEs waiting to be scheduled and the number of connected UEs in the serving cell. For the low-load region with less than 6 UEs (3XR+3eMBB), the number of queued UEs is similar across all schedulers. However, notable improvements are observed with the proposed scheduler under high-load conditions with more than 9 UEs (6XR+3eMBB). This is due to PDU-set segmentation and latency awareness, resulting in more efficient transmission of PDU-sets and a reduction in queued UEs. Compared to WPF, the proposed scheduler shows a decrease in the average number of queued UEs from about 7 UEs/cell to 5 UEs/cell in the load of 11 UEs. 

\begin{figure}
        \centering
        \includegraphics[width=0.37\textwidth,trim={0 0 0 0},clip] {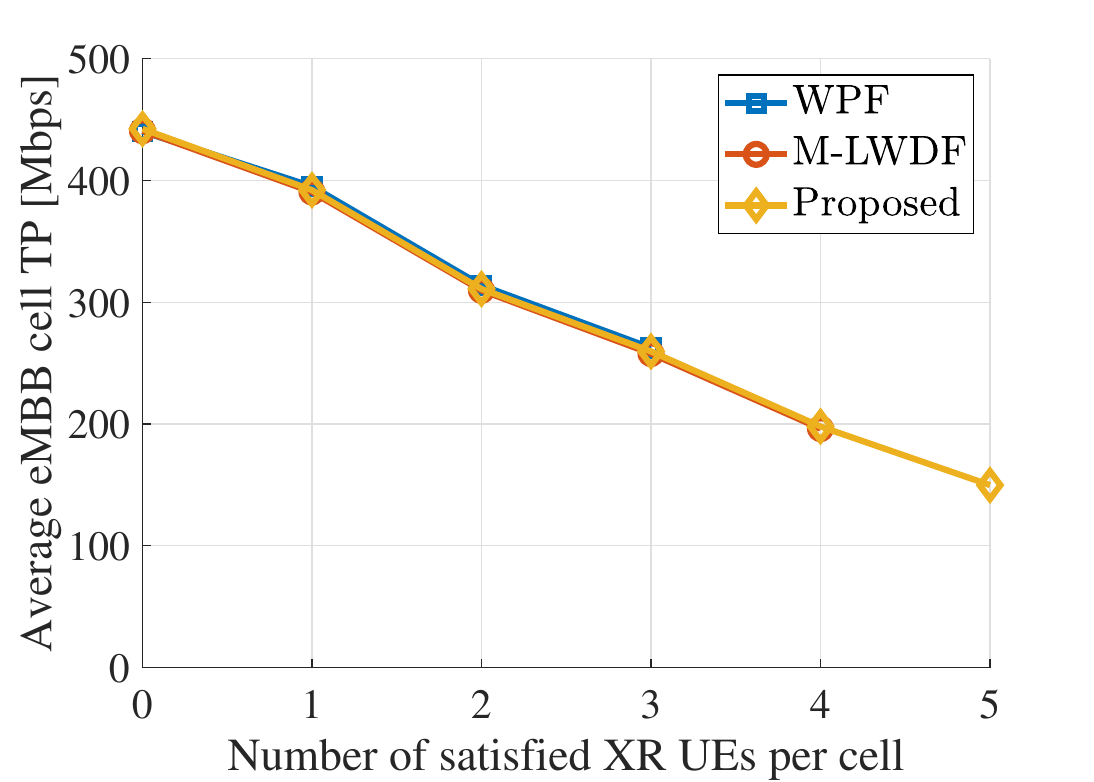} 
                 
        \caption{Average eMBB cell TP versus number of connected XR UEs per cell for three different schedulers.}
        \label{P5:Fig:eMBB_Cell_TP}
    \end{figure}

    Figure \ref{P5:Fig:eMBB_Cell_TP} shows the impact of adding XR UEs on the average eMBB cell TP for different schedulers. This is built on the average cell TP samples obtained from the eMBB UEs per simulation run. 
    As observed from Figure \ref{P5:Fig:eMBB_Cell_TP}, the eMBB TP performance declines linearly as a function of the number of XR users per cell. Using a simple minimum mean square error fit reveals that the eMBB throughput declines with $1.3 \times N \times 45$ Mbps, where $N$ is the number of XR users (XR UE’s source data rate is 45 Mbps). This confirms that there is a cost for the eMBB performance of serving XR traffic with strict QoS requirements.
    Furthermore, when comparing the proposed scheduler with the baseline schedulers, negligible differences are observed. Overall, our scheduler not only offers a notable increase in XR capacity but also keeps the same eMBB TP compared to the baseline schedulers. This means that our scheduler efficiently uses radio resources to serve XR UEs, while still allocating a fair amount of resources to eMBB UEs.

\section{Conclusions}\label{P5:Conclusions}
    In this paper, we studied a dynamic packet scheduling algorithm to improve XR capacity in a multi-cell multiuser multi-service 5G-Advanced network. 
    The proposed scheduler considers the PDU-set information such as the fraction of the remaining delay budget as well as the fraction of PDU-set transmitted bits. In particular, leveraging the awareness of PDU-set segmentation significantly improves scheduling decisions since the scheduler prioritizes XR UEs with smaller remaining PSDB and smaller buffered data over other XR UEs. 
    The performance of the proposed solution is evaluated with advanced dynamic system-level simulations. Results show that the proposed scheduler achieves up to a 45\% improvement in XR capacity while maintaining the same eMBB cell TP achieved with the baseline schedulers.


\end{document}